\documentstyle[aps,prb]{revtex}

\newcommand{\beq}{\begin{equation}}
\newcommand{\eeq}{\end{equation}}

\begin{document}

\twocolumn[\hsize\textwidth\columnwidth\hsize\csname
@twocolumnfalse\endcsname

\title{Slabs of stabilized jellium: Quantum-size and self-compression effects}

\author{I. Sarria,$^{1}$ C. Henriques,$^{2, 3}$ C. Fiolhais,$^{3}$ and J. M.
Pitarke$^{1,4}$}

\address{ $^1$ Materia Kondentsatuaren Fisika Saila, Zientzi Fakultatea, Euskal
Herriko Unibertsitatea, 644 Posta kutxatila,\\ 48080 Bilbo, Basque Country,
Spain\\
$^2$ Departamento de F\'{\i}sica, Faculdade de Ci\^encias e
Tecnologia, Universidade Nova de Lisboa,\\ P-2825-114 Caparica, Portugal\\
$^3$ Center for Computational Physics, Department of Physics, University of
Coimbra,\\ P-3004-516 Coimbra, Portugal\\
$^4$ Donostia International Physics Center (DIPC) and Centro Mixto
CSIC-UPV/EHU,\\
Donostia, Basque Country, Spain}

\date{\today}

\maketitle

\begin{abstract}
We examine thin films of two simple metals (aluminum and lithium) in the
stabilized jellium model, a
modification of the regular jellium model in which a constant potential is
added
inside the metal to stabilize the system for a given background density. We
investigate quantum-size effects on the surface energy and the work
function. For a given film thickness we also evaluate the density yielding
energy
stability, which is found to be slightly higher than the equilibrium density of
the bulk system and to approach this value in the limit of thick slabs. A
comparison of our self-consistent calculations with the predictions of the
liquid-drop model shows the validity of this model.
\end{abstract}

\pacs{PACS numbers: 71.15.Mb, 71.15.Nc, 71.45.Gm}
]

\section{Introduction}

Thin films or slabs are systems made out of a few layers of atoms: they are
finite in one direction and infinite in the other two perpendicular directions.
Various theoretical models are available to calculate the electronic structure
of slabs. One of the important features coming out of these calculations is the
so-called quantum-size effect (QSE),\cite{schu} i.e., the influence of the
finite
size on various physical properties of the slab. These effects, which can
be experimentally recognized,\cite{jak} decrease as the size of the slab
increases. In fact, surface energies and work functions of the semi-infinite
system are often derived from thin-slab calculations, which are simply
extrapolated to this limit.\cite{boet5,boet6,fiore,sabin}

The simplest model to predict the electronic structure of simple
$sp$-bonded metals is the jellium model,
where the ions are replaced by a positive neutralizing background. Within this
model, the QSE of thin films was examined by Schulte.\cite{schu} He found an
oscillatory behavior of the work function, as a function of the thickness of the
slab. The same oscillatory behavior is found for the surface energy, defined as
the energy required, per unit area of new surface formed, to split the solid in
two along a plane.\cite{Lang}

The jellium model has been referred to as giving insight into the realistic QSE
appearing in real systems.\cite{feibel,vicen,wojci} Notwithstanding
important differences, an oscillatory pattern also appears in atomistic
first-principles slab calculations of both the work function and the surface
energy. Although the presence of the lattice may obscure the periodicity and the
amplitude of the QSE, extrema were found at positions which agree with the
jellium results. However, in the case of first-principles calculations
difficulties arise (due to the cumbersome numerics) when one is to
extract well-converged surface properties from thin films made of typically 2
to 15 layers.\cite{fiore,sabin} Hence, the clean jellium QSE, with
no uncertainties in the extrapolated results, remains as a guide for more
realistic investigations.

In this paper, we consider slabs in the framework of a simple modification of
the jellium model which yields energy stability against changes in the
background density. This so-called stabilized jellium \cite{Kiej} or
structureless
pseudopotential model yields realistic results, especially in the case of
metals with high valence-electron
density. For instance, the stabilized jellium model predicts positive
surface energies that increase rapidly at high electron densities, as
shown by experiment, while the jellium model predicts surface energies that
are strongly negative at these densities. The stabilized jellium model, first
introduced by Perdew, Tran, and Smith\cite{per} and similar to the ideal-metal
concept developed by Shore and Rose,\cite{shore,rose} has been applied to the
study of surfaces\cite{car,kie} and clusters.\cite{bra2} In a
way, the stabilized jellium is in  between the jellium model and more
sophisticated atomistic approaches: although it is still a continuous model (one
may choose slabs with arbitrary thickness) with an analytical expression for the
bulk energy, its
physical predictions are in reasonable agreement with experiment. Besides
including electrostatic corrections to the jellium model, the stabilized jellium
model contains an averaged pseudopotential correction.

We calculate the self-consistent energetics (surface energy and work
function) of slabs of stabilized jellium, with use of the local-density
approximation (LDA) of density-functional theory (DFT).\cite{ks,dre}
We take two metals: Al ($r_{s}=2.07$, $Z=3$) and Li ($r_{s}=3.24$,
$Z=1$), investigate the QSE, and compare our self-consistent slab calculations
with those obtained for a semi-infinite stabilized jellium. We also test an
extrapolation rule,\cite{pit2} which has already been used to
describe  non-local surface energies of the bounded electron
gas.\cite{pit}

Although the stabilized jellium model can be tailored to give face-dependent
results,\cite{per,kie,per3} it cannot describe the inhomogeneous relaxation
predicted by first-principles calculations where the distances between atomic
planes of the same family are optimized. However, an interesting effect
displayed
by the stabilized jellium model, which cannot be accounted for by the jellium
model, is the so-called self-compression \cite{per2} (or self-expansion, in the
case of charged systems \cite{vie,bra}) of clusters. This effect, which can be
classically viewed as the compression of a finite system due to the surface
tension, is most prominent for small systems and almost negligible in the case
of large clusters.

We investigate here the self-compression of thin films. We fix the size of
the system along the direction perpendicular to the surface, and search for the
background density which minimizes the total energy per valence electron of the
slab. The equilibrium density is found to increase as the thickness of the slab
decreases, and to converge to the bulk electron density in the
infinite-thickness limit. Furthermore, the equilibrium electron-density parameter
$r_s^*$ is found to oscillate with the slab thickness, as a manifestation of
the QSE, but the general trend is found to be well described within the
liquid-drop model (LDM)\cite{mye,perl} based only on the knowledge of the bulk
energy per unit volume and the surface energy. We discuss the relationship
between this self-compression effect and the relaxation of metal slabs predicted
by atomistic first-principles calculations.

In section II we present briefly the stabilized jellium model for slabs. In
section III we discuss the results we obtained within this model. The main
conclusions are drawn in section IV, where further comments on the relationship
between the stabilized jellium and more elaborated models are made.
Equations are written in atomic units throughout, i.e., $e^2=\hbar=m_e=1$.

\section {Slabs of stabilized jellium}

The stabilized jellium model\cite{per} takes into account the lattice ions, but
keeps the essential simplicity of the jellium model. The total energy is
obtained as a functional of the electron density $n({\bf r})$, in the following
way:
\begin{eqnarray}\label{efu}
E_{SJ}\lbrack{n,n_+}\rbrack=E_{J}\lbrack{n,n_+}\rbrack+
\left(e_{M}+\bar{w}_{R}\right)\int d^3{r}\,{n_+({\bf r})}\cr
+{\langle\delta v\rangle}_{WS}\int d^3{r}\,{n_+({\bf r})\over\bar{n}}
\,\left[n({\bf r})-n_+({\bf r})\right],
\end{eqnarray}
where
\begin{equation}
n_+=\bar n\,\Theta({\bf r})
\end{equation}
represents a positive neutralizing background density, $\Theta({\bf r})$
being a function which equals 1 inside a given surface and 0
outside, and
\beq\label{dens}
\bar n={{3}\over{4 \pi {r_{s}}^3}}
\eeq
is the average valence-electron density. $E_{J}$ is the regular-jellium
total energy,
$e_{M}$ is the Madelung energy arising from the Coulomb interaction between
a uniform negative
background inside the spherical Wigner-Seitz cell and a point ion at its
center,
\beq\label{em}
e_{M}=-{9\,Z^{2/3}\over 10\,r_s},
\eeq $\bar {w}_{R}$ is the average value of the
repulsive non-Coulomb part of the Ashcroft empty-core pseudopotential,
\beq\label{wr}
\bar{w}_{R}=2\pi\bar{n}{r_{c}^2},
\eeq
and $\langle\delta
v\rangle_{WS}$ represents the difference between the local pseudopotential and
the jellium potential, averaged over the Wigner-Seitz cell,
\beq\label{deltav}
\langle\delta v\rangle_{WS}={3\,r_c^2\over 2\,r_s^3}-{3\,Z^{2/3}\over
10\,r_s}.
\eeq
The core radius $r_c$ of the Ashcroft empty-core pseudopotential is chosen to
stabilize the metal for given values of the electron-density parameter $r_s$ and
the chemical valence $Z$.

The two terms added to the regular-jellium energy $E_{J}$ are a volume term
and a surface term. They simply account for the subtraction of the
spurious self-interaction of the positive jellium background and the inclusion
of a constant structureless potential inside the metal. This procedure may be
understood as a first-order perturbation to a jellium system, but with the
perturbation treated in an averaged manner.

The density functional of Eq. (\ref{efu}) represents the total energy of an
arbitrary inhomogeneous system. In the case of an infinite uniform
system, the
equilibrium density is obtained from the bulk stability condition
\beq\label{ebulk}
{d\epsilon_{SJ}^{bulk}\over dr_s}=0,
\eeq
where
\begin{equation}\label{bulk}
\epsilon_{SJ}^{bulk}=e_J^{bulk}+e_M+\bar w_R
\end{equation}
represents the average bulk energy per valence electron, $e_J$ being the
regular-jellium contribution. Within this model any individual metal minimizes
the energy at a given equilibrium density, while the jellium energy presents a
single minimum at $r_s\sim 4.2$ close to the electron-density parameter of
sodium.

We consider slabs of stabilized jellium. Slabs are translationally invariant in
the plane of the surface, which is assumed to be perpendicular to the $z$ axis.
Hence, the single-particle wave functions can be separated into a plane wave
along the surface and a component $\phi(z)$ describing motion normal to the
surface with energy $\epsilon$. This component is obtained by solving
self-consistently the Kohn-Sham equation
\beq\label{ko}
\left[-{1\over 2}{d^2\over dz^2}+V_H(z)+V_{xc}(z)+V_{ps}(z)\right]
{\phi}(z)=\epsilon\,\phi(z),
\eeq
where $V_H(z)$ represents the Hartree electrostatic potential, $V_{xc}(z)$ is the
exchange-correlation potential, and $V_{ps}(z)$ accounts for the
pseudopotential,
\beq\label{vdelta}
V_{ps}(z)={\langle\delta v\rangle}_{WS}\,\Theta(z).
\eeq
$V_{xc}(z)$ is obtained in the LDA, using the
electron-gas correlation
energy of Ceperley and Alder,\cite{Ceperley} as parameterized by Perdew and
Wang.\cite{perw} Essentially the same results are obtained from the
parameterizations of Vosko, Wilk, and Nusair\cite{vos} and of Perdew and
Zunger.\cite{perz} We have not chosen to use extensions such as the
generalized gradient approximation (GGA),\cite{perg} since the LDA has been
shown to give surprisingly good results in describing the properties of
jellium planar surfaces.\cite{kur}

Outside the positive background the electron-density profile $n(z)$ decays
rapidly from its bulk value $\overline n$. The electronic system can
therefore be
taken to be finite in the $z$ direction by assuming that $n(z)$ actually
vanishes at a given distance $z_0$ from the surface. Hence, we introduce
infinite potential walls at a distance $z_0$ from each surface, and follow
Ref.\onlinecite{Eguiluz} to expand the wave functions $\phi(z)$ in a Fourier
sine series. The distance $z_0$ (typically 2 or 3 Fermi wavelengths) and the
number of sines kept in the expansion of the wave functions $\phi(z)$ have
been chosen to be sufficiently large for our calculations to be insensitive
to the precise values employed. These calculations have been compared with
others which we have carried out for a semi-infinite electron system by
using the Monnier-Perdew code\cite{mon} for the numerical integration of Eq.
(\ref{ko}).

For a given thickness $L$ of the slab, we obtain the surface energy from the
difference between the total energy of Eq. (\ref{efu}) and the
corresponding
result for a homogeneous electron gas of density $n_+$, i.e.,
\beq\label{se}
\sigma(L)={1\over 2A}\left[E_{SJ}(L)-{\bar
n}\,L\,A\,\epsilon_{SJ}^{bulk}\right],
\eeq
where A is the normalization area. The work function is obtained as the
difference between the computed values for the vacuum and Fermi levels of our
electron system.

\section{Results and discussion}

First of all, we compare jellium and stabilized-jellium electron densities
$n(z)$ and effective potentials,
\begin{equation}
V_{eff}(z)=V_H(z)+V_{xc}(z)+V_{ps}(z).
\end{equation}
Jellium and stabilized-jellium valence-electron densities and effective
potentials for an Al
slab of $L=2\,\lambda_F$
[$\lambda_F=\left(32\,\pi^2/9\right)^{1/3}r_s$ is the Fermi wavelength]
are shown in Fig. 1, together with the positive
background density $n_+$. We note that the stabilized-jellium electron
density is steeper at the two surfaces, so that the electronic spill-out is
slightly smaller within this model. This is due to the fact that electrons feel
a deeper effective potential. Both jellium and stabilized-jellium electron
densities exhibit quantum oscillations inside the metal, the so-called Friedel
oscillations,\cite{Lang} and an exponential decay outside.

Figs. 2 and 3 show our calculated stabilized-jellium surface energies for slabs
of Al and Li, respectively, as obtained from Eq. (\ref{se}) versus the thickness
$L$ of the slab. Both curves show damped oscillations with minima occurring at
the slab width $L\sim n\,\lambda_F/2$ ($n=1, 2,...$). The same QSE, which
reflects the quantization of the electronic motion along one direction, is known
to occur within the jellium model.\cite{schu}

Both the average bulk energy per valence electron $\epsilon_{SJ}^{bulk}$ and
the surface energy of the semi-infinite stabilized jellium
\beq\label{sigma}
\sigma=\lim_{L\to \infty}\sigma(L)
\eeq
may be obtained from a linear fit of the following equation:
\begin{equation}
{E_{SJ}(L)\over A}=2\,\sigma+\bar n\,L\,\epsilon_{SJ}^{bulk},
\end{equation}
where $E_{SJ}(L)$ represents the total energy of Eq. (\ref{efu}). Following this
procedure, we reproduce the bulk energy of Eq. (\ref{bulk}) and predict surface
energies of $925 \,{\rm erg/cm}^2$ and
$311\,{\rm erg/cm}^2$ for Al and Li, respectively. These surface energies,
represented in Figs. 2 and 3 by horizontal solid lines, agree with those reported
in Ref.\onlinecite{kie} for semi-infinite media.

An alternative procedure to extrapolate the surface energy $\sigma$ of the
semi-infinite medium from our calculated thin-film surface energies
$\sigma(L)$ is to use the relation\cite{pit2}
\begin{equation}\label{limit}
\sigma={\sigma(L_n-\lambda_F/4)+\sigma(L_n)+\sigma(L_n+\lambda_F/4)\over 3},
\end{equation}
where $L_n$ represents the threshold width for which the $n$th subband for the
$z$ motion is first occupied. Analytical insight for this procedure is
encountered within the infinite-barrier model (IBM), where the effective
potential $V_{eff}(z)$ is replaced by an infinite square well and the
one-particle wave functions $\phi(z)$ are simply sines. Based on this
procedure, the numerical error introduced in $\sigma$ by our slab calculations
is found to be within $0.1\%$. The advantage of this algorithm is that we
simply need three points to obtain the asymptotic limit, while the linear
fitting may yield erroneous results if one only takes a few thin
films.

Slabs with $L<0.5\,\lambda_{F}$ are interesting in their own, since they can be
constructed in the laboratory, e.g., by joining two different semiconductors.
Nevertheless, we do not give results for these ultra-thin slabs, since they fall
within the two-dimensional limit where the three-dimensional LDA and
GGA formulae for exchange and correlation are known to fail.\cite{poll}

For comparison, first-principles thin-film calculations of the surface
energy of the most dense faces of Al and Li [(111) for fcc Al and (0001) for hcp
Li] are represented in Figs. 2 and 3 by solid circles and triangles, with the
slab width of a $\nu$-layer unrelaxed crystalline film taken to be $\nu$ times
the interplanar distance. For Al there is reasonable agreement between our
stabilized-jellium results and atomistic first-principles calculations, the
amplitude of the stabilized-jellium oscillations being comparable to that
exhibited by first-principles calculations. For Li, however, there is a serious
discrepancy between stabilized-jellium and first-principles calculations. Since
lithium has been found to behave to some extent like a covalent solid rather
than a free-electron gas,\cite{doll,boetr,birken2,kokko} it is not expected to
be well described by a jellium-like model.

A face-dependent approach extension of the stabilized-jellium model consists in
obtaining the self-consistent electron density by adding to the constant
potential ${\langle\delta v\rangle}_{WS}$ a structure-dependent corrugation
factor.\cite{per,kie} This procedure yields an increased surface
energy [horizontal dashed-dotted lines of Figs. 2 and 3], which in the case of Al
is found to be close to the experimental result.

Figs. 4 and 5 exhibit our calculated stabilized-jellium work functions
for slabs of Al and Li, respectively, as a function of the thickness $L$ of the
slab, together with first-principles thin-film calculations. As in
the case of the surface energy, a procedure similar to that of Eq.
(\ref{limit}) yields a work function [represented by horizontal solid lines] that
agrees within less than $0.1\%$ with the result we also obtain after solving Eq.
(\ref{ko}) for the semi-infinite medium, a precision that is difficult to
achieve by a fitting procedure. For $L\sim 0.5\lambda_F$, the QSE yields
oscillations with relative amplitudes of $\sim 20\%$ and $\sim10\%$ for Al
and Li, respectively. For Al both the amplitude and the oscillation pattern are
comparable to those exhibited by atomistic calculations. In the case of a
3-layer film of Al(111), the slab width is $L\sim 4\,(\lambda_F/2)$.
Hence, the stabilized-jellium model predicts a minimum for this film, which is
in reasonable agreement with the deep minimum exhibited by atomistic
calculations with $\nu=3$. In the case of Li(0001), the stabilized-jellium
model predicts a minimum for a one-layer film [$L\sim 1\,(\lambda_F/2)$], also
in agreement with the minimum exhibited by first-principles calculations with
$\nu=1$. Finally, we note that adding a structure-dependent corrugation factor
to the slabilized-jellium ${\langle\delta v\rangle}_{WS}$ constant potential
yields a smaller value of the work function [horizontal dashed-dotted lines of
Figs. 4 and 5], which in the case of Al is in reasonable agreement with the
experiment. For Li, both the stabilized-jellium model and first-principles
calculations predict work functions that are well above the experimental
result.

For given values of the equilibrium-density parameter $r_s$ and the
valence $Z$, all these calculations have been carried out with the core radius
$r_c$ [characteristic of each metal] that is obtained from the bulk stability
condition expressed by Eq. (\ref{ebulk}). However, while at the equilibrium
density $\bar n$ of Eq. (\ref{dens}) the infinite homogeneous system is stable,
at this density a finite system is not stable against changes of the background
density, i.e.,
\beq\label{dern}
{d(E/N)\over dr_s}\neq 0,
\eeq
where $N$ represents the particle number. Instead, there is a modified
equilibrium-density parameter $r_s^*$,
which stabilizes the finite system. This modified parameter depends on the size
$L$ of our system and is expected to approach $r_s$ as $L\to\infty$.

Fig. 6 shows the result of our full self-consistent Kohn-Sham calculations of
the deviation $r_s^*-r_s$, as a function of the thickness $L$ of the slab. These
calculations indicate that there is a self-compression effect, which is more
pronounced when the two surfaces are separated by a multiple of
$\sim\lambda_{F}/2$.

The self-compression effect exhibited in Fig. 6 may be approximately predicted
with use of the LDM, a simple model to evaluate the total energy of a finite
system.\cite{mye,perl} In this model, the energy is the sum of a volume term
(the bulk energy per unit volume, $\bar{n}\,\epsilon_{bulk}$, times the volume)
and a surface term (the surface energy, $\sigma$, times the transversal area):
\beq\label{ldm}
E_{LDM}=\bar{n}\epsilon_{bulk}\,V+\sigma\,A.
\eeq
For fixed $r_c$, and evaluated at the {\it bulk} equilibrium-density parameter
$r_s$, 
\beq\label{der}
{d(E_{LDM}/N)\over dr_s}={A\over N}\,{d\sigma\over dr_s}+\sigma\,{d(A/N)\over
dr_s}>0.
\eeq
The first term is positive, as can be found from the
data in Table I of Ref.\onlinecite{per2}. For a fixed slab width $L$, the
second term is also positive, and the surface term self-compresses, therefore,
stabilized-jellium slabs. The deviation of the electron density parameter
$r_s^*$ obtained from the LDM stability condition
\beq\label{der}
{d(E_{LDM}/N)\over dr_s}=0
\eeq
with respect to the {\it bulk} equilibrium density parameter $r_s$ is also
plotted in Fig. 6, showing that the LDM provides a nice average of our
self-consistent Kohn-Sham calculations, as previously demonstrated in the case of
clusters.\cite{per2} 

In Ref. \onlinecite{vicen}, thin films of Be with 1-3 layers were examined and a
jellium version of a crystalline calculation was considered. The
electron density parameter $r_s^*$ needed to define each slab was derived from
the optimized (relaxed) structural parameters. The results reported in
Ref.\onlinecite{vicen} are in agreement with the compression effect we
report here, with $r_s^*$ increasing with the number of layers and approaching
the {\it bulk} equilibrium-density parameter $r_s$ as $L\to\infty$. These results
show deviations of the electron density parameter, $r_s^*-r_s$ of $\sim 3.2\%$,
$1.9\%$ and $0.9\%$ for thin films with 1, 2 and 3 layers, respectively. This is
in agreement with our stabilized-jellium calculations, which in the case of
thin films with $\sim 2$ layers of Li and Al predict (see Fig. 6) differences
between $r_s^*$ and $r_s$ of $\sim 1.6\%$ and $1.8\%$, respectively. The
self-compression of structural parameters in ultra-thin crystalline films has
also been discussed in terms of the so-called coordination model
which, however, seems to fail in some cases (see,
e.g., Ref.\onlinecite{boetrmd}).

Finally, we note that if for each value of $L$ the corresponding
equilibrium-density parameter $r_{s}^*$ is taken, instead of the {\it bulk}
parameter $r_{s}$, modified surface energies and work functions are obtained
which are quite similar to those displayed in Figs. 2-5. This is in contrast
with the discussion of Ref.\onlinecite{vicen}.

\section{Conclusions}

We have modeled thin films of two simple metals, aluminum and lithium,
using the stabilized-jellium model, and have studied the convergence of some
physical quantities (work function and surface energy) to the
semi-infinite planar-surface results. We have found the same oscillatory
behavior which is typical of the QSE in jellium. Although this
behavior also shows up in atomistic first-principles thin-film calculations, the
clean QSE of continuous background models is obscured in the more realistic
calculations. A trend consisting in surface energy
minima coinciding with work
function maxima was reported for first-principles crystalline calculations.
\cite{bat,kie2} However, within the stabilized-jellium model we have found minima
and maxima of both quantities at the same positions [as also reported in Ref.
\onlinecite{boet6} from first-principles for Al(111)]. On the other hand, we
have found that both the absolute and the relative amplitude of
stabilized-jellium QSE oscillations are larger for aluminum than for lithium,
in agreement with first-principles evaluations. The disagreement between our
stabilized-jellium results for lithium and the more realistic atomistic {\it
all-electron} calculations cannot be attributed to some property of the
pseudopotential, and simply shows that this metal does not display free-electron
behaviour.

Stabilized-jellium slabs of aluminum and lithium have been found not to be stable
at the {\it bulk} equilibrium density, the size-dependent equilibrium density
being larger. This self-compression effect, which was already known to exist for
clusters, has been found to become more important as the slab width decreases.
Both LDM and full self-consistent DFT calculations have shown a larger
self-compression for aluminum than for lithium, which is a consequence of the
larger surface energy of the former material. The self-compression of thin
simple-metal films is a general rule that is also exhibited by
atomistic first-principles calculations, where the unitary cell of thin
films is found to be slightly smaller than that of the bulk solid.

The stabilized jellium model is computationally as simple as the jellium
model; however, for the two high-density metals we have considered, it is much
more realistic. In particular, we have found it to be more realistic for aluminum
than for lithium. The stabilized-jellium model is adequate to obtain general
qualitative conclusions and an understanding of trends of simple metals but,
obviously, is unable to provide precise quantitative conclusions on particular
metals. These can only be extracted from the now standard first-principles, but
computationally more demanding, calculations.

\section{Acknowledgments}

The authors gratefully acknowledge J. P. Perdew for fruitful discussions. This
project has been supported by the University of the Basque Country, the Basque
Hezkuntza, Unibertsitate eta Ikerketa Saila, the Spanish Ministerio de
Educaci\'on y Cultura, and the Portuguese PRAXIS XXI Program (Project
PRAXIS/2/2.1/FIS/473/94).

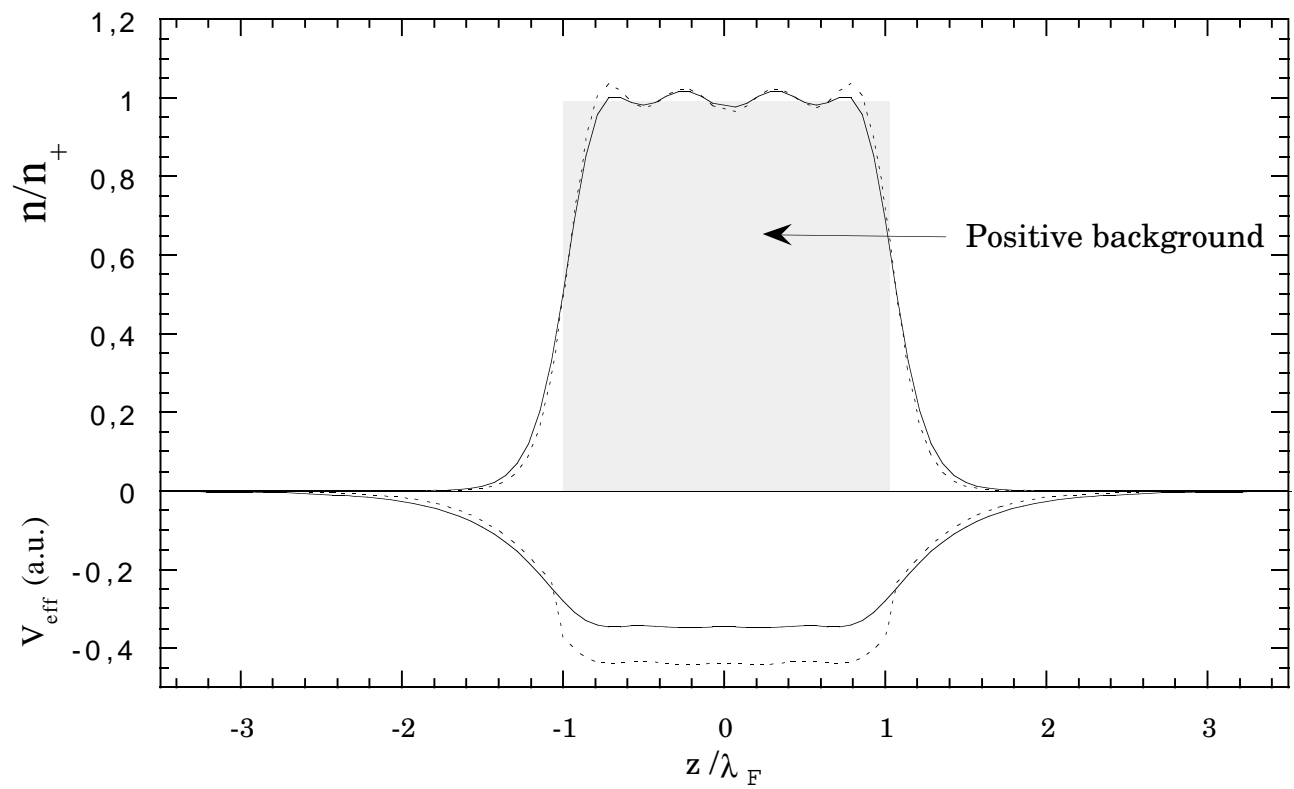
\begin{figure}
\caption{Normalized valence-electron density in the jellium
model (solid line) and in the stabilized-jellium model (dashed line) for a slab
of Al ($r_s=2.07$) with thickness $L=2\,\lambda_F$. The background density
is represented by the dark area. The figure also displays the effective
potential $V_{eff}(z)$ in each model (solid line for the jellium model and dashed
line for the stabilized-jellium model).}
\end{figure}

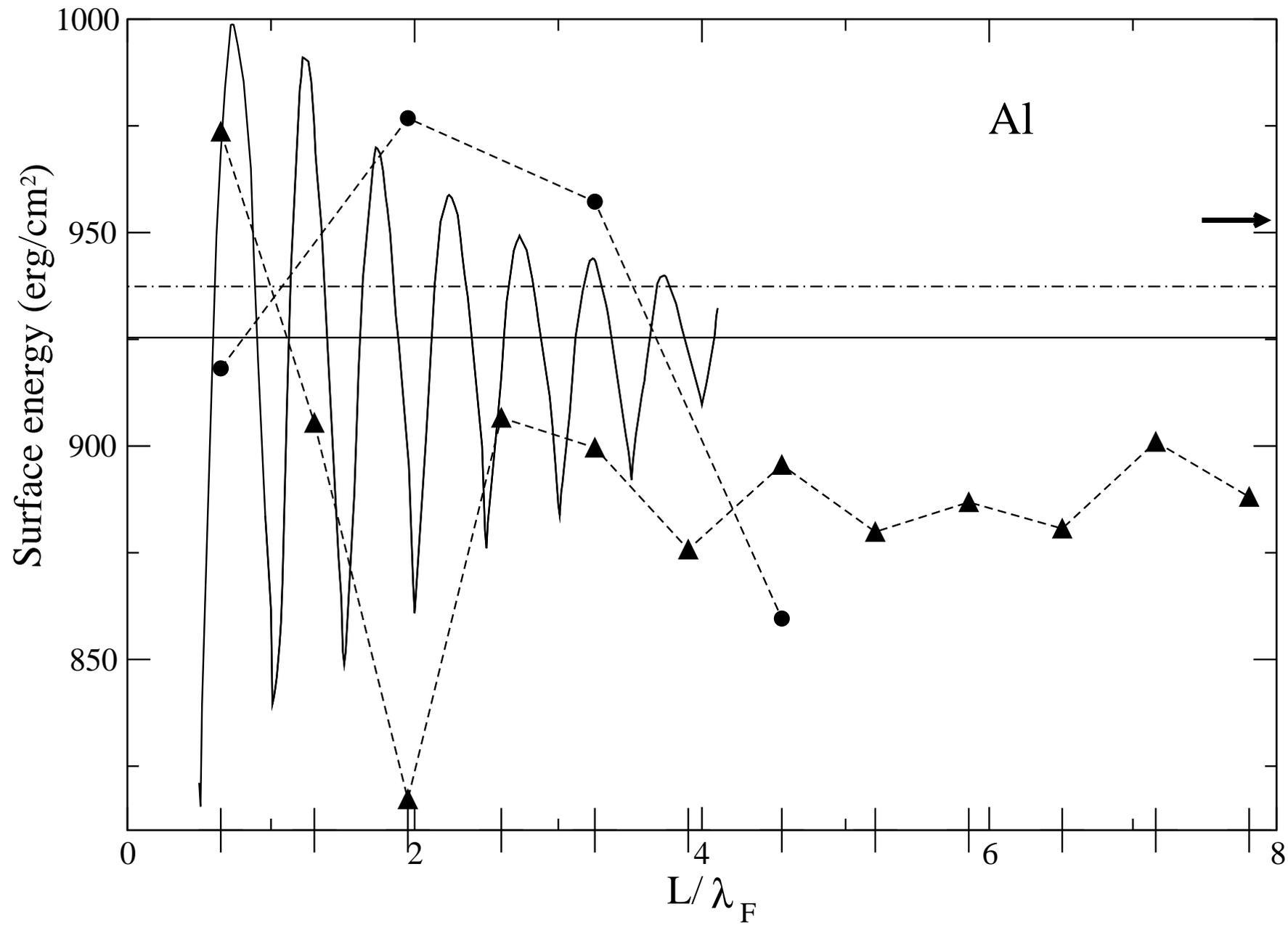
\begin{figure}
\caption{Surface energy and QSE in aluminum ($r_s=2.07$). Large vertical marks
across the horizontal axis show the widths of unrelaxed fcc Al(111) slabs with
$\nu=1,..., 12$ atomic planes. The width $L$ is given by
$L=\nu\,(\protect\sqrt 3\protect/3)\,a$, $a$ being the lattice
parameter $a=(16\,\pi\,Z/3)^{1/3}r_s$. The
solid oscillating line shows our calculated surface energy of flat
stabilized-jellium slabs. Solid and dashed-dotted horizontal lines represent
our calculated surface energy of semi-infinite flat Al (solid line) and fcc
Al(111) (dashed-dotted line) stabilized jellia. The zero-temperature
extrapolation of the experimental liquid-metal surface tension of
Ref.\protect\onlinecite{tyson}\protect\, divided by
$1.2$,\protect\cite{perl}\protect\, is represented by an horizontal arrow. For
comparison, atomistic first-principles calculations from
Refs.\protect\onlinecite{bat}\protect\, and
\protect\onlinecite{boet6}\protect\, are also displayed, by solid circles and
triangles, respectively. The surface energies of
Ref.\protect\onlinecite{bat}\protect\, were obtained using the self-consistent
pseudopotential method combined with an independent calculation of the bulk
energy per electron. The surface energies of
Ref.\protect\onlinecite{boet6}\protect\, were obtained within an all-electron
scheme with the use of a linear-combination-of-gaussian-type-orbitals fitting
function (LCGTO-FF) and with the bulk energy per electron extracted from the
slab calculations. Dashed lines are to guide the eye.}
\end{figure}

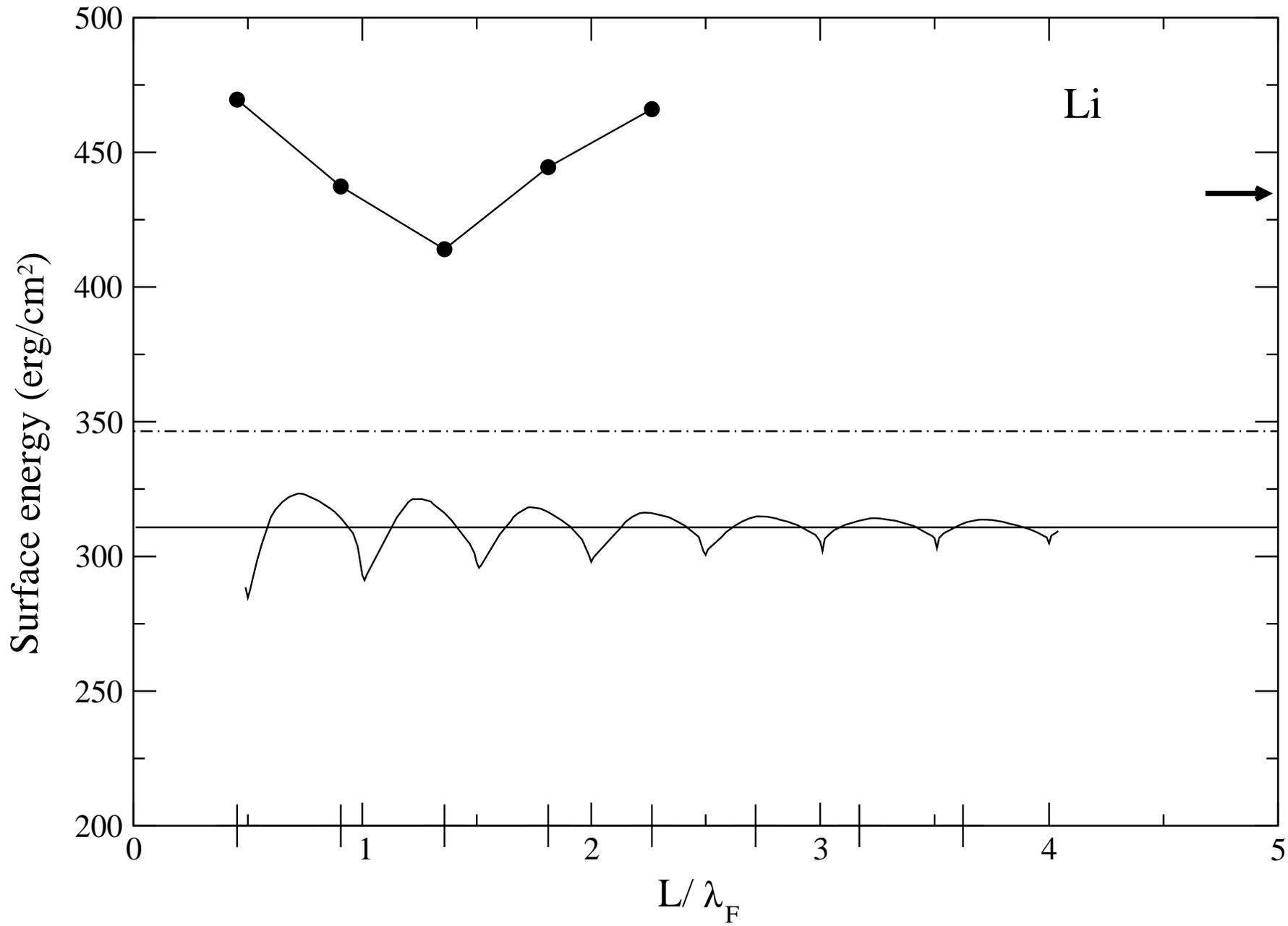
\begin{figure}
\caption{Surface energy and QSE in lithium ($r_s=3.24$). Large vertical marks
across the horizontal axis show the widths of unrelaxed hcp Li(0001) slabs
with $\nu=1,..., 12$ atomic planes [$c/a=1.64$
\protect\cite{boetr}\protect, which corresponds to
$r_s=3.13$]. The slab width is $L=\nu\,a/2$
and the structural-parameters' ratio
$c/a=\left(16\,\protect\sqrt{3}\protect\,\pi\,Z/9\right)\,\left(r_s/a\right)^3$.
The solid oscillating line shows our calculated surface energy of flat
stabilized-jellium slabs. Solid and dashed-dotted horizontal lines represent
our calculated surface energy of semi-infinite flat Li (solid line) an hcp
Li(0001) (dashed-dotted line) stabilized jellia. The horizontal arrow has the
same meaning as in Fig. 2. For comparison, atomistic all-electron calculations
from Refs.\protect\onlinecite{boetr}\protect\, and
\protect\onlinecite{birken2}\protect\, are also displayed, by solid circles.
These surface energies were obtained with the use of a LCGTO-FF and with the 
bulk energy per electron extracted from the slab calculations. Dashed lines are
to guide the eye.}
\end{figure}

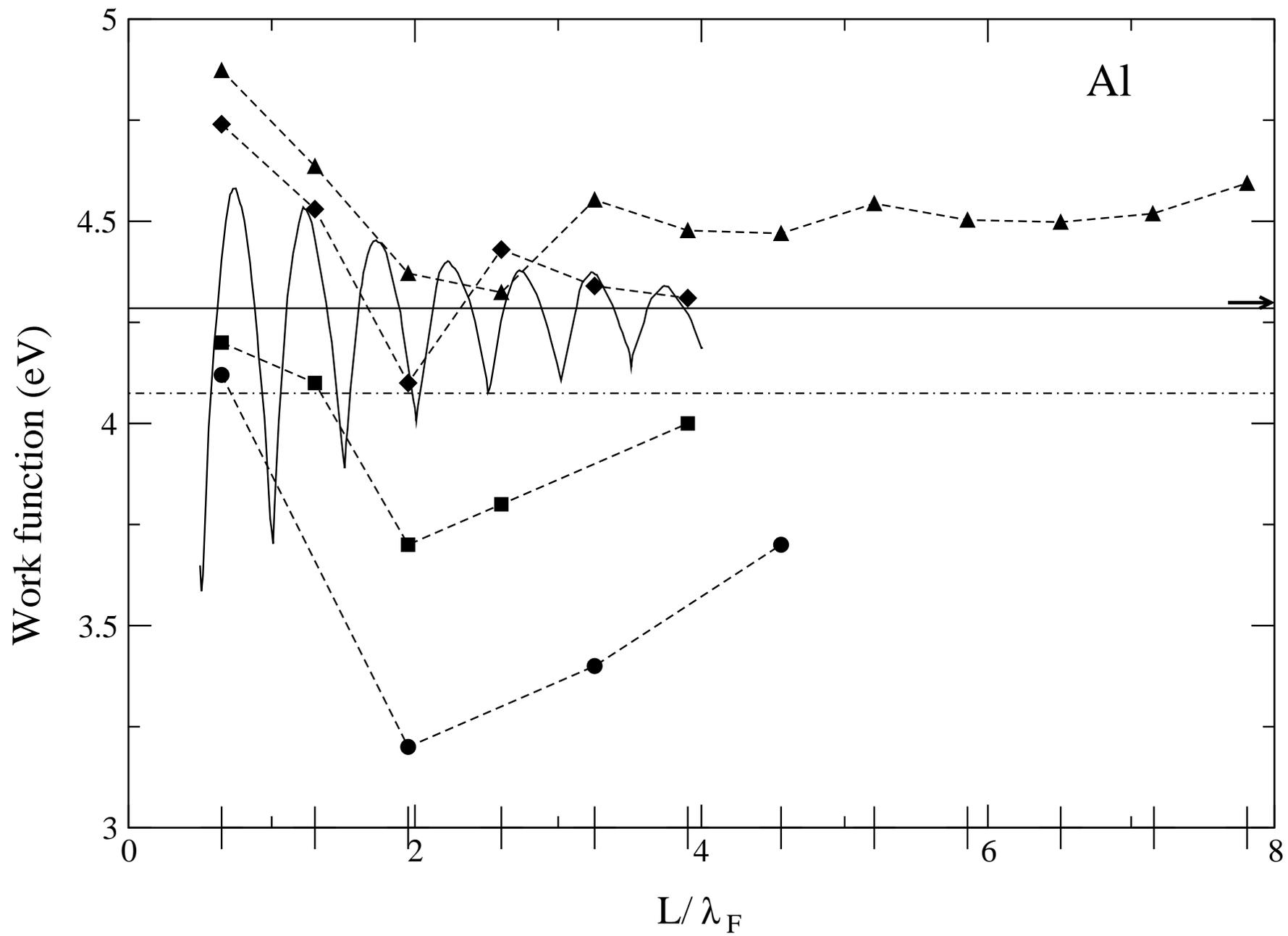
\begin{figure}
\caption{Work function and QSE in aluminum ($r_s=2.07$). All symbols have the
same meaning as in Fig. 2. For comparison, atomistic all-electron calculations
from Refs.
\protect\onlinecite{feibel}\protect\, and \protect\onlinecite{feibel2}\protect\, are also displayed, by solid squares
and rhombs, respectively. The work functions of
Ref.\protect\onlinecite{feibel}\protect\, were obtained within the LCAO scheme,
and those of Ref.\protect\onlinecite{feibel2}\protect\, were obtained with the
use of surface linearized augmented plane waves (SLAPW). The experimental
polycrystalline work function of Ref.\protect\onlinecite{michael}\protect\,
is represented by an horizontal arrow.}
\end{figure}

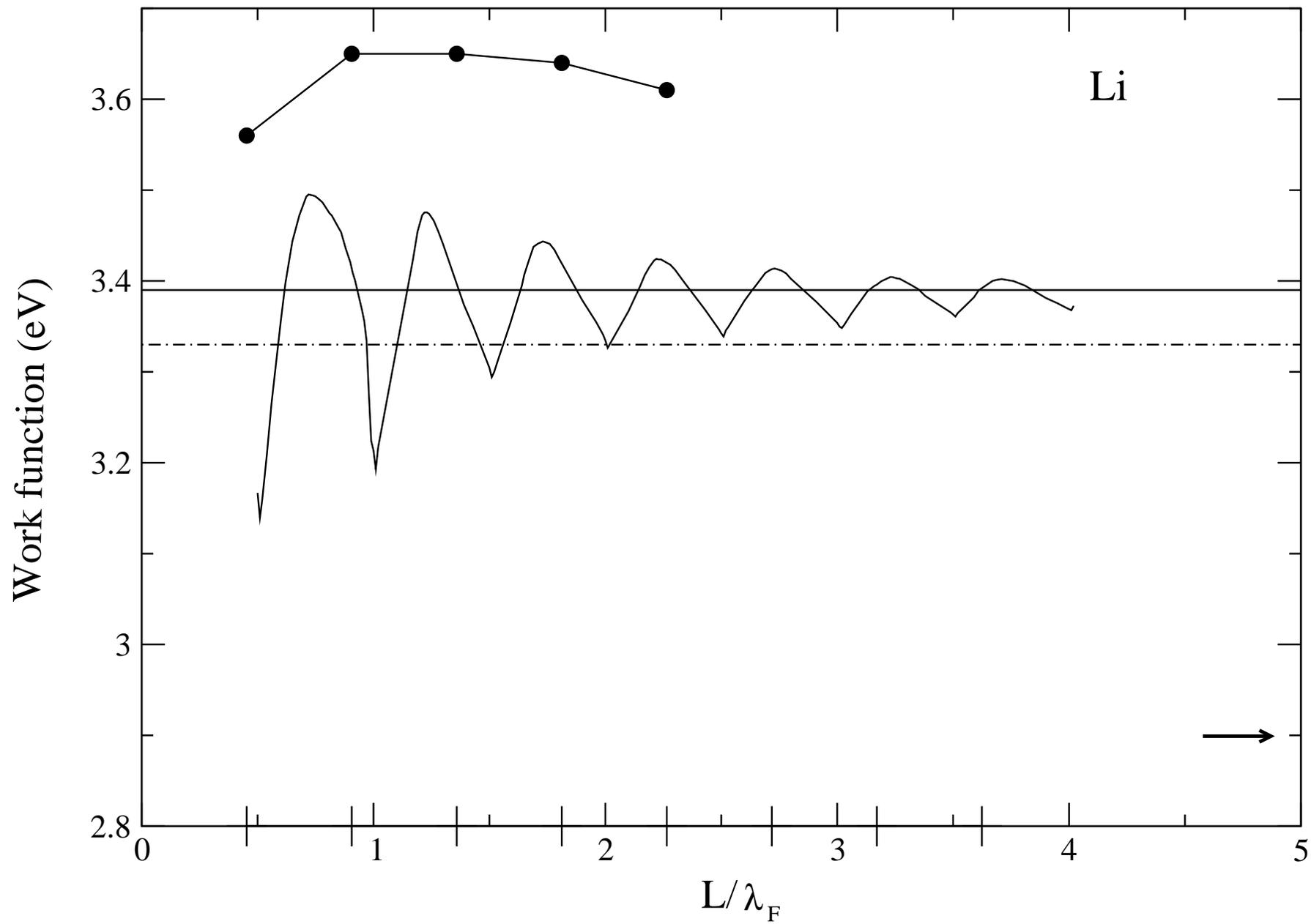
\begin{figure}
\caption{Work function and QSE in lithium ($r_s=3.24$). All symbols have the same
meaning as in Fig. 3. The experimental polycrystalline work function of
Ref.\protect\onlinecite{michael}\protect\, is represented by an horizontal
arrow.}
\end{figure}

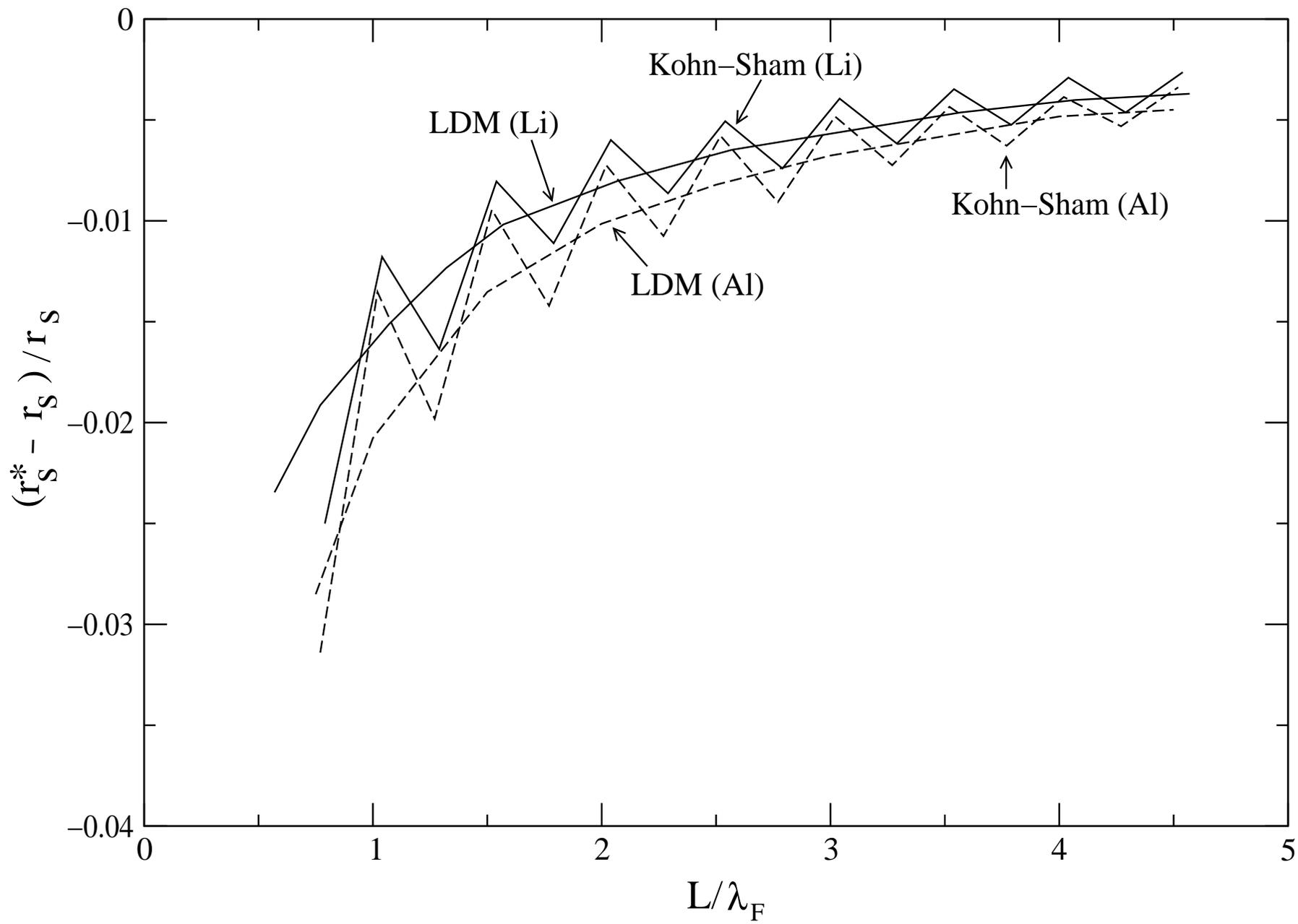
\begin{figure}
\caption{Relative difference between the actual equilibrium-density parameter
$r_{s}^*$ and the {\it bulk} density parameter $r_s$ for aluminum (dashed lines)
and lithium (solid lines) stabilized-jellium films, as a function of the slab
width $L$.}
\end{figure}


\begin{references}
\bibitem{schu} F. K. Schulte, Surf. Sci. {\bf 55}, 427 (1976).
\bibitem{jak} R. C. Jaklevic and J. Lambe, Phys. Rev. B {\bf 12}, 4146 (1975).
\bibitem{boet5} J. C. Boettger, Phys. Rev. B {\bf 49}, 16798 (1994).
\bibitem{boet6} J. C. Boettger, Phys. Rev. B {\bf 53}, 13133 (1996).
\bibitem{fiore} V. Fiorentini and M. Methfessel, J. Phys.: Condens. Mat. {\bf
8}, 6525 (1996).
\bibitem{sabin} J. C. Boettger, J.R. Smith, U.
Birkenheuer, N. R\"osch, S. B. Trickey, J. R. Sabin,
and S. P. Apell, J. Phys.: Condens. Mat. {\bf 10}, 893 (1998).
\bibitem{Lang} See, e.g., N. D. Lang, Solid State Phys. {\bf 28}, 225 (1973).
\bibitem{feibel} P.J. Feibelman, Phys. Rev. B {\bf 27}, 1991 (1983).
\bibitem{vicen} J. L. Vicente, A. Paola, A. Razzitte, E. E. Mola, and
S. B. Trickey, Phys. Stat. Sol. B {\bf 155}, K93 (1989).
\bibitem{wojci} K. F. Wojciechowski and H. Bogdanow, Surf. Sci. {\bf 397},
53 (1998).
\bibitem{Kiej} A. Kiejna, Prog. Surf. Sci. {\bf 61}, 65 (1999).
\bibitem{per} J. P. Perdew, H. Q. Tran, and E. D. Smith, Phys. Rev. B {\bf 42},
11627 (1990).
\bibitem{shore} H. B. Shore and J. H. Rose, Phys. Rev. Lett. {\bf 66}, 2519
(1991); J. H. Rose and H. B. Shore, Phys. Rev. B {\bf 43}, 11605 (1991).
\bibitem{rose} H. B. Shore and J. H. Rose, Phys. Rev. B {\bf 59}, 10485 (1999).
\bibitem{car} C. Fiolhais and J. P. Perdew, Phys. Rev. B {\bf 45}, 6207 (1992).
6207.
\bibitem{kie} A. Kiejna, Phys. Rev. B {\bf 47}, 7361 (1993). Small
differences between surface energies reported in Ref.\onlinecite{car} and those
of this reference are due to improved numerics in the later.
\bibitem{bra2} M. Brajczewska, C. Fiolhais, and J. P. Perdew, Int. J. Quantum
Chem. {\bf 27}, 249 (1993).
\bibitem{ks} P. Hohenberg and W. Kohn, Phys. Rev. {\bf 136}, B864 (1964); W.
Kohn and L. J. Sham, Phys. Rev. {\bf 140}, A11333 (1965).
\bibitem{dre} R. M. Dreizler and E. K. U. Gross, {\it Density Functional Theory}
(Springer-Verlag, Berlin, 1990).
\bibitem{pit2} J. M. Pitarke and A. G. Eguiluz (unpublished).
\bibitem{pit} J. M. Pitarke and A. G. Eguiluz, Phys. Rev. B {\bf 57}, 6329
(1998).
\bibitem{per3} J. P. Perdew, Progr. Surf. Sci. {\bf 48}, 795 (1995).
\bibitem{per2} J. P. Perdew, M. Brajczewska, and C. Fiolhais, Sol. Stat. Com.
{\bf 88}, 795 (1993).
\bibitem{vie} A. Vieira, M. Brajczewska, C. Fiolhais, and J. P. Perdew, Int. J.
of Quantum Chem. {\bf 60}, 1537 (1996).
\bibitem{bra} M. Brajczewska, A. Vieira, C. Fiolhais, and J. P. Perdew, Prog.
Surf. Sci. {\bf 53}, 315 (1996).
\bibitem{mye} W. D. Myers and W. J. Swiatecki, Ann. Phys. (N. Y.) {\bf 55}, 395
(1969); {\bf 84}, 186 (1974).
\bibitem{perl} J. P. Perdew, Y. Wang, and E. Engel, Phys. Rev. Lett. {\bf 66},
508 (1991).
\bibitem{Ceperley} D. M. Ceperley and B. J. Alder, Phys. Rev. Lett. {\bf 45},
1196 (1980).
\bibitem{perw} J. P. Perdew and Y. Wang, Phys. Rev. B {\bf 45}, 13244 (1992).
\bibitem{vos} H. Vosko, L. Wilk, and M. Nusair, Can. J. Phys. {\bf 58}, 1200
(1980).
\bibitem{perz} J. P. Perdew and A. Zunger, Phys. Rev. B {\bf 23}, 5048 (1981).
\bibitem{perg} J. P. Perdew, K. Burke, and M. Ernzerhof, Phys. Rev. Lett. {\bf
77}, 3865 (1996).
\bibitem{kur} S. Kurth and J. P. Perdew, Phys. Rev. B {\bf 59}, 10461 (1999);
Z. Yan, J. P. Perdew, S. Kurth, C. Fiolhais, and L. Almeida, Phys. Rev. B
{\bf 61}, 2595 (2000).
\bibitem{Eguiluz} A. G. Eguiluz, D. A Campbell, A. A. Maradudin, and
R. F. Wallis, Phys. Rev. B {\bf 30}, 5449 (1984); A. G. Eguiluz, Phys. Scr.
{\bf 36}, 651 (1987).
\bibitem{mon} R. Monnier and J. P. Perdew, Phys. Rev. B {\bf 17}, 2595 (1978).
\bibitem{poll} L. Pollack and J. P. Perdew, J. Phys.: Condens. Mat. {\bf
12}, 1241 (2000).
\bibitem{doll} K. Doll, N. M. Harrison, and V. R. Saunders, J. Phys.:
Condens. Mat. {\bf 11}, 5007 (1999).
\bibitem{boetr} J. C. Boettger and S. B. Trickey, Phys. Rev. B {\bf 45}, 1363
(1992).
\bibitem{birken2} U. Birkenheuer, J.C. Boettger, and N. R\"{o}sch, Surf. Sci.
{\bf 341}, 103 (1995).
\bibitem{kokko} K. Kokko, P.T. Salo, R. Laihia, K. Mansikka, Surf. Sci. {\bf
348}, 168 (1996).
\bibitem{boetrmd} J. C. Boettger, S. B. Trickey, F. Muller-Plathe, G. H. F.
Diercksen, J. Phys.: Condens. Mat. {\bf 2}, 9589 (1990).
\bibitem{bat} I. P. Batra, S. Ciraci, G. P. Srivastava, J. S. Nelson, and C. Y.
Fong, Phys. Rev. B {\bf 34}, 8246 (1986).
\bibitem{kie2} A. Kiejna, J. Peisert, and P. Scharoch, Surf. Sci. {\bf 432}, 54
(1999).
\bibitem{tyson} W. R. Tyson and W. A. Miller, Surf. Sci. {\bf 62}, 267 (1977).
\bibitem{feibel2} P. J. Feibelman and D. R. Hamann, Phys. Rev. B {\bf 29}, 6463
(1984).
\bibitem{michael} H. B. Michaelson, J. Appl. Phys. {\bf 48}, 4729 (1997).

\end{references}
\end{document}